\documentclass[3p,twocolumn]{elsarticle}

\usepackage{amssymb}
\usepackage{amsmath}
\usepackage{xcolor}
\usepackage{bm}
\newcommand{\Lop}{\mathsf{L}}
\newcommand{\Bop}{\mathsf{B}}
\newcommand{\Dop}{\mathsf{D}}

\usepackage{float}
\usepackage{hyperref}
\hypersetup{
    colorlinks=true,
    linkcolor=blue,
    filecolor=blue,
    urlcolor=blue,
    citecolor=blue,
    }
\journal{Ocean Modelling}

\begin{document}

\begin{frontmatter}



\title{Discrete variance decay analysis of spurious mixing}

\author[jacobs,awi]{Tridib Banerjee}
\author[jacobs,awi]{Sergey Danilov}
\author[iow]{Knut Klingbeil}

\affiliation[jacobs]{organization={Dept of Mathematics, Constructor University},
            city={Bremen},
            postcode={28759},
            country={Germany}}
\affiliation[awi]{organization={Dept of Climate Sciences \& Dynamics, Alfred Wegener Institute for Polar and Marine Research},
            city={Bremerhaven},
            postcode={27570},
            country={Germany}}
\affiliation[iow]{organization={Dept of Physical Oceanography \& Instrumentation, Leibniz Institute for
Baltic Sea Research Warnemunde (IOW)},
            city={Rostock},
            postcode={18119},
            country={Germany}}

\begin{abstract}
Expressions for local discrete variance decay (DVD) rates are directly derived from discrete tracer equations without any assumptions on discrete fluxes of the second moment. Spurious mixing (SM) associated with numerical implementations of scalar advection and diffusion is thus estimated. The new framework is shown to avoid the need for second-moment flux definition when solved on finite-volume cell edges but still invoke certain second-moment fluxes when the DVD rates are partitioned to participating cell nodes. These implied discrete fluxes are shown to differ from those proposed in earlier literature (but share the same dissipative part) and thus reveal the non-uniqueness of their nature. They are shown to be ambiguous for high-order advection schemes introducing uncertainty to the locality of any estimates produced by a DVD approach. Additional damping of flux divergence through temporal averaging or some coarse-graining is thus shown to be necessary. Through the application of this technique, SM is found to be correlated with the distribution of eddy kinetic energy. The contribution from vertical advection to SM is found to be relatively small and correlated with the distribution of buoyancy fluxes. The explored high-order schemes are found to demonstrate levels of spurious mixing which may locally exceed background physical mixing. 
\end{abstract}

\begin{highlights}
\item Closed-form expressions for Discrete Variance Decay rates (DVD) are derived.
\item Error in DVD estimates stemming from unfiltered flux divergence are revealed.
\item Spurious Mixing (SM) is found to follow distribution of eddy kinetic energy.
\item SM due to vertical advection is found to follow the distribution of buoyancy fluxes.
\item Advection schemes are found to produce SM comparable to background physical mixing.
\end{highlights}

\end{frontmatter}


\section{Introduction}
\label{intro}
High-order upwind schemes commonly used for horizontal advection in ocean circulation models may mix watermass properties (\citet{gerdes1991influence}). Part of this mixing may simulate the effect of isopycnal mixing from missing eddies if meshes are coarse, but the other part is purely numerical and is needed to eliminate perturbations with wrong phase and group velocities that are created by the schemes at grid scales (\citet{weaver1997numerical}). It also contributes to spurious diapycnal mixing (SDM) due to misalignment between layer (\citet{marchesiello2009spurious}) and isopycnal surfaces or through spurious isopycnal mixing and cabbeling (\citet{marsh2000cabbeling}, \citet{mcdougall1987thermobaricity}). This numerical mixing is often an uncontrolled (and undesirable) source of mixing in ocean modelling. Recent literature has thus proposed several approaches for its diagnosis and mitigation.\\

The most widely used diagnostic relies on reference potential energy (RPE), which is the minimum potential energy obtained by adiabatically sorting density, following \citet{winters1995available}. By tracking the growth of RPE as a function of time, one can judge about SDM or estimate an effective diapycnal diffusivity as demonstrated by \citet{griffies2000spurious}. After the study by \citet{Ilicak2012}, which linked the amount of SDM to eddy kinetic energy, the RPE-based diagnostics is routinely used in ocean circulation models (see, e.g., \citet{Petersen2015,karna2018,Mohammadi-Aragh2015,Gibson2017}). It provides a global measure of SDM that includes contributions from all processes. \citet{Gibson2017} generalized the RPE based approach to separate between contributions from horizontal and vertical advection, which is possible in their case because the vertical advection is implemented as a separate remapping step. \citet{ilicak2016quantifying} proposes further generalization in the RPE method, which is an extension of the available potential density method by \citet{winters2013available} and allows one to get the spatial distribution of the RPE growth.\\

Approaches used by \citet{getzlaff2010diagnostics,getzlaff2012diagnostics,hill2012controlling} are essentially based on releasing a patch of passive tracer and tracking its temporal evolution. A broader class of approaches rely on the watermass transformation framework of \citet{walin1982relation}. \citet{megann2018estimating}, following the earlier work by \citet{lee2002spurious}, computed the meridional overturning streamfunction in the density framework. The diapycnal velocities in this framework allow one to estimate the effective diapycnal diffusivity and its spurious part by subtracting physical diapycnal (vertical) diffusivity. A similar approach is used in \citet{megannztilde} to assess the performance of $\tilde{z}$ vertical coordinates. Other variants of the framework of \citet{walin1982relation} were adopted to a particular tracer such as temperature (\citet{holmes2019diathermal},  \citet{holmes2021geography}) or salinity ( \citet{li2018transformation}, \citet{maccready2018estuarine}, \citet{burchard2019}, \citet{burchard2020}, \citet{wang2021quantification}, \citet{schlichting2023quantification}), allowing one to estimate total and numerical mixing of the respective tracer.\\

Methods relying on discrete variance decay (DVD) approach the same problem of estimating spurious mixing from a numerical side. One tries to assess the destruction of variance of a particular tracer (or velocity) considering the numerical implementation of the transport (advection--diffusion) equation. This can be done in each cell of the computational mesh and at each time moment, and is fully consistent with the numerical approximations of a model. However, the change of the variance in a cell differs from the DVD rate by the divergence of second moment fluxes (see detailed explanation below), and the estimate of these fluxes presents the main uncertainty for assessing {\em local} DVD rates. In some cases, like for the first-order upwind scheme, the interpretation is straightforward (see \citet{maqueda2006second} and \citet{burchard2008comparative}). \citet{klingbeil2014quantification} present one possible general derivation of fluxes of second moment based on the decomposition and recombination of subvolumes inside single grid cells.\\

This work revisits the computation of local DVD rates. It departs from equation for the first moment (i.e. temperature or salinity) and tries to find expressions for the DVD rates directly. This is possible if one changes the viewpoint: from cells to cell faces, looking for changes in the second moment due to flux of the first moment. Implied fluxes of second moment are computed thereafter and compared with those proposed in \citet{klingbeil2014quantification}. They are shown to disagree even in the limit of small time steps and remain ambiguous for high-order methods (like third-order upwind advective schemes) generally, with the implication that numerical estimates of DVD rates may still contain contributions from flux divergences. This does not influence the global variance decay, only its local distribution which thus needs certain coarse-graining to reduce the effect of remaining flux divergence. In our direct approach, the net DVD rate is obtained as a combination of separate contributions from diffusion and advection with the contributions also being separable in horizontal and vertical directions. Vertical diffusion is considered as the physical part as it results from physically grounded parameterizations. The rest (horizontal diffusion and diffusion build in the advection scheme) is referred to as numerical mixing. We note that some authors refer to explicit horizontal diffusion as physical, but here only the part that relies on physically-grounded parameterizations is called `physical'.      \\

This approach does not distinguish between diapycnal and isopycnal directions. However, in setups with a linear monovariate equation of state the net DVD rate is by construction purely diapycnal. Such setups can be used to learn about spurious diapycnal mixing associated with a particular advection scheme, and/or its dependence on flow regimes. The advantage of this DVD concept lies in its ability to provide local estimates, with reservations mentioned above. The test case of \citet{soufflet2016effective} is used to examine several third- and fourth-order schemes available in FESOM2, revealing that the associated DVD rate correlates with the distribution of eddy kinetic energy and can exceed the DVD rate of physical mixing locally.\\

The paper is structured as follows. It first derives independent closed-form expressions for DVD rates due to scalar transport - diffusion and advection in section \ref{sec:theory}. It also explores the uncertainty associated with any \textcolor{blue}{local} DVD analysis. This is then followed by the experimental section \ref{sec:exp} which tests various aspects and dependencies of the proposed DVD formulation - fluxes, decomposition, spatial and temporal resolutions, and effects of time-stepping. The section also applies the method proposed in this paper to various advection schemes over a range of upwinding parameter and compares their spurious mixing. Sections \ref{sec:disc} and \ref{sec:conc} summarises the findings and comments on future possibilities.

\section{Theory}
\label{sec:theory}
\subsection{Preliminaries}
Consider the advection--diffusion equation for a tracer $T$,
\begin{equation}
  \partial_tT+\nabla\cdot(\mathbf{v}T-\bm{\kappa}\nabla T)=0
\label{eq:3DT}
\end{equation}
in some domain $D$. Solid boundaries are impermeable and `insulated'. There is no surface flux of $T$, $\mathbf{v}=(\mathbf{u},w)$ is the full velocity, with $\mathbf{u}$ the horizontal velocity, $w$ the vertical velocity, $\nabla\cdot\mathbf{v}=0$, and $\bm{\kappa}$ the 3 by 3 symmetric positive diffusivity matrix. We will assume that it is diagonal for simplicity. Equation (\ref{eq:3DT}) multiplied by $2T$ gives the equation for second moment,
\begin{equation}
  \partial_tT^2+\nabla\cdot(\mathbf{v}T^2-2T\bm{\kappa}\nabla T)=-2\nabla T\cdot\bm{\kappa}\nabla T=-\chi
\label{eq:3DT2}
\end{equation}
If integrated over the domain, the flux term drops out giving,
$$
\partial_t\langle T^2\rangle_D=-\langle \chi\rangle_D
$$
where $\langle A\rangle_D$ denotes averaging over domain $D$. Since $\langle T\rangle_D=const$, the second moment can be replaced by the variance $\langle T^2\rangle_D-\langle T\rangle_D^2$ and the equation above becomes the variance decay equation, where $\langle \chi\rangle_D$ is the global variance decay rate. In analogy, $\chi$ in equation (\ref{eq:3DT2}) is referred to as a local variance decay rate. It can be determined given the rate of change of second moment and the divergence of flux of second moment. In the continuous case of equation (\ref{eq:3DT2}) the variance decay is due to diffusion only. In the discrete case, numerical implementation of advection may contain diffusive truncation errors, or such errors are introduced by limiters imposed to maintain positivity or the TVD (total variation diminishing) property. The discrete variance decay (DVD) rate therefore is contributed by both diffusion and advection. If a discrete analog of (\ref{eq:3DT2}) were available, the DVD rate could be estimated if discretized fluxes of second moment were available.
However, in the discrete case one discretizes (\ref{eq:3DT}), and not (\ref{eq:3DT2}). One may propose a plausible definition for fluxes of $T^2$, which is the approach of \citet{klingbeil2014quantification}. In the latter work, the discrete flux of second moment is computed as,
$$
\mathbf{F}=-2T^{n+1}\bm{\kappa}_i\nabla T^{n+1}- 2T^{n}\bm{\kappa}_e\nabla T^{n}+\mathbf{v}\tilde{T}^2
$$
where the first two terms are the implicit and explicit parts of diffusion, and $\tilde{T}$ is the interfacial value of tracer in the discrete version of tracer equation (\ref{eq:3DT}). We propose a different approach by starting from the discrete version of (\ref{eq:3DT}) and looking at the change in second moment created by fluxes of first moment through cell faces.
\subsection{Vertical Diffusion}
Consider an example of 1D implicit diffusion, which corresponds to vertical diffusion in ocean global circulation models,
\begin{equation}
\partial_t T=\partial_x K \partial_x T
\label{eq:1DTc}
\end{equation}
with related second-moment equation,
\begin{equation}
\partial_t T^2+\partial_x(-K\partial_x T^2)=-2K(\partial_x T)^2
\label{eq:1DT2c}
\end{equation}
Implicit discretization of (\ref{eq:1DTc}) on a uniform mesh is written as,
\begin{equation}
\begin{split}
    T^{n+1}_i-T^n_i&=-(\Delta t/\Delta x)(f_{i+1/2}-f_{i-1/2})\\
    \quad f_{i+1/2}&=-(K/\Delta x)(T_{i+1}^{n+1}-T_i^{n+1})
\end{split}
\label{eq:1DT}
\end{equation}
Here, $i$ is the control volume index and $n$ is the time step index. The fluxes $f_{i+1/2}$ are defined at faces between the control volumes. We are going to derive the balance of second moment relying solely on (\ref{eq:1DT}). First, this equation is multiplied with $2T^*_i=(T^{n+1}_i+T^n_i)$. Its left hand side (LHS) becomes the difference of squares, i.e. the discretized time derivative of the second moment,
\begin{equation}
    \begin{split}
&(T^{n+1}_i)^2-(T^n_i)^2\\
&=-(\Delta t/\Delta x)(2T^*_if_{i+1/2}-2T^*_if_{i-1/2})
    \end{split}
    \label{eq:1DT2}
\end{equation}
The right hand side (RHS) of this equation should be interpreted as the combination of discrete divergence of fluxes of second moment and the DVD rate taken with opposite sign. The second moment flux is unknown, but we can proceed without knowing it by re-grouping the terms on the RHS. Consider the face $i+1/2$. Flux $f_{i+1/2}$ contributes to the tendency of second moment in cell $i$ with $-2f_{i+1/2}T_i^*$. It also contributes with $2f_{i+1/2}T_{i+1}^*$ to the cell $i+1$. The net contribution associated with flux $f_{i+1/2}$ is,
$$
\tilde{\chi}_{i+1/2}=\chi_{i+1/2}\Delta x=-2 f_{i+1/2}(T_{i+1}^{*}-T_i^{*})
$$
with,
\begin{equation}\label{eq: Dver equation}
    \begin{split}
        \chi_{i+1/2}
&= 2K \frac{T_{i+1}^{n+1}-T_i^{n+1}}{\Delta x}\frac{T_{i+1}^{*}-T_i^{*}}{\Delta x}\\
&= -2 f_{i+1/2}\frac{T_{i+1}^{*}-T_i^{*}}{\Delta x}
    \end{split}
\end{equation}
This is the expression for DVD rate. The tilde value includes `volume'. Re-grouping removes the contribution from fluxes and leaves the sink. In fact, it is equivalent to considering all cells $i$ together because $\tilde{\chi}_{i+1/2}$ is related to two cells. Note also that equation (\ref{eq:1DT2}) cannot be transformed to the form of (\ref{eq:1DT2c}) locally because the discrete analog of $2K(\partial_xT)^2$ is defined at faces and not cells, i.e. lies in a different discrete space. The two contributions $-2f_{i+1/2}T_i^*$ and $2f_{i+1/2}T_{i+1}^*$ related to flux $f_{i+1/2}$ differ in sign and absolute value. They can be thought of as coming from a flux of second moment and a sink,
$$
-2f_{i+1/2}T_i^*=-F_{i+1/2}-\alpha\tilde{\chi}_{i+1/2}
$$
and,
$$
2f_{i+1/2}T_{i+1}^*=F_{i+1/2}-(1-\alpha)\tilde{\chi}_{i+1/2}
$$
where $\alpha$ is a partitioning parameter. The selection of $\alpha$ is arbitrary, and we take $\alpha=1/2$ to treat the cells in the same way. This selection simultaneously defines the flux,
\begin{equation}
 F_{i+1/2}=(T^*_i+T^*_{i+1})f_{i+1/2}
\label{eq:vdf}
\end{equation}
and equal partitioning of $\tilde{\chi}_{i+1/2}$ between cells $i$ and $i+1$. Equation (\ref{eq:1DT2}) takes the form,
\begin{equation}
    \begin{split}
        &\Delta x\frac{(T^{n+1}_i)^2-(T^n_i)^2}{\Delta t} + F_{i+1/2}-F_{i-1/2}\\
        &= - \frac{1}{2}( \tilde{\chi}_{i-1/2} + \tilde{\chi}_{i+1/2} )
    \end{split}
\end{equation}
The flux (\ref{eq:vdf}) differs from its respective contribution in \citet{klingbeil2014quantification} by the use of starred tracer related to time moment $n+1/2$ instead of time moment $n+1$. They will agree in the limit of small time steps. Time stepping effects might be significant if the respective CFL number is high. Even though the flux is ambiguous, it only distributes the DVD rate $\tilde{\chi}_{i+1/2}$ between cells $i$ and $i+1$. The DVD rate averaged over several cells  will not be affected much by the precise form of distribution (the selection of factor $\alpha$ above).\\

This ambiguity can be avoided by working with face values but this is not always convenient in practice. Also, $\chi_{i+1/2}$ in (\ref{eq: Dver equation}) is an approximation to the decay rate of second moment $\chi=2K(\partial_x T)^2$ around $i+1/2$ except for the lack of full synchronicity in time. The quantity $\tilde{\chi}_{i+1/2}$ can be associated with the interval between centers of $i$ and $i+1$ cells. Similar manipulations can be done for horizontal diffusion and advection as explained further. It will be seen however, that there are more ambiguities in the definitions of DVD rate and second moment fluxes when high-order operators are used.
\subsection{Horizontal diffusion}
This section uses the discretization of FESOM which assumes that tracer $T$ is placed at vertices and is associated with the median-dual control volumes around them. The reason is the specific form of biharmonic operator used. A single index $v$ enumerates vertices, and we talk about mesh edges that connect vertices. Otherwise, the expressions below are general and valid for other meshes if edges are replaced by lines connecting cell centers. The discrete harmonic diffusion operator is written as,
\begin{equation}
  (\Lop T)_v=(1/A_v)\sum_{n\in V(v)}(T_n-T_v)K_{nv}
\end{equation}
Here, $V(v)$ is the set of neighbour vertices connected to $v$ by edges and $A_v$ is the area of the scalar control volume around $v$. An edge is identified by the pair of vertex indices. The boundary of control volumes consists of segments connecting mid-edges to triangle centers (two such segments are associated with each edge). $K_{nv}$ above is diffusivity related to the edge modified with geometrical factors. The biharmonic operator is implemented in two steps. First, we compute,
\begin{equation}
(\Dop T)_v=\sum_{n\in V(v)}(T_{n}-T_v)K^{1/2}_{nv}
\end{equation}
Note the absence of numerical factor related to the area. Also note that because of this absence, $K_{nv}$ has the dimension of m$^2$/s, as the harmonic diffusivity. Second, we compute,
\begin{equation}
(\Bop T)_v=-(1/A_v)\sum_{n\in V(v)}[(\Dop T)_{n}-(\Dop T)_v]K^{1/2}_{nv}
\end{equation}
Note the minus sign because of negative eigenvalue.
\subsubsection{Harmonic diffusion}
Control volume around vertex $v$ contributes to the variance tendency with
\begin{equation}
2T^*_v\sum_{v'\in V(v)}(T^n_{v'}-T^n_v)K_{v'v}
\end{equation}
Therefore, contribution from the face associated to edge $v'v$ into total (area integrated) DVD is,
\begin{equation}
\tilde{\chi}_{v'v}=2(T^*_v-T^*_{v'})(T^n_{v'}-T^n_v)K_{v'v}
\end{equation}
If time step is sufficiently small, this contribution is negative definite. We split it into two equal parts and place them into control volumes $v$ and $v'$. Note that we are dealing with area-weighted contributions, i.e. division by $A_v$ will be required afterwards to get $\chi$ in units of tracer squared per second.

\subsubsection{Biharmonic diffusion}
For a biharmonic operator, contribution to the area-weighted DVD rate from edge $vv'$ will be,
\begin{equation}\label{eq:BH equation}
\tilde{\chi}_{vv'}=-2(T^*_v-T^*_{v'})\left[(\Dop T)^n_{v'}-(\Dop T)^n_v\right]K_{v'v}
\end{equation}
In contrast to the harmonic operator, this expression is not sign definite, even for small time steps. We still can split the edge contributions equally between control volumes around $v$ and $v'$ as in the case above, obtaining $\tilde{\chi}_v=\sum_{v'\in V(v)}(1/2)\tilde{\chi}_{vv'}$, but the result may contain contributions that can still be seen as flux divergence. To eliminate such flux divergences we first sum all $\chi_{vv'}$ over edges $vv'$ and re-arrange the sum as a sum over vertices $v$ (the rationale being that $\Dop T$ is defined at the same locations as $T$ while the gradients are defined at faces in the case of harmonic operator). The total area integrated variance decay, $\sum_v\tilde{\chi}_{v}$ is also the sum over edges of the edge contributions, i.e,
\begin{equation}
 \sum_{vv'}\tilde{\chi}_{vv'}=\sum_v\tilde{\chi}_v=(1/2)\sum_v\sum_{v'\in V(v)}\tilde{\chi}_{vv'}
\end{equation}
Here 1/2 is because of equal partitioning between control volumes. This leads to,
\begin{equation}
    \begin{split}
    \sum_v\tilde{\chi}_{v}&=-2\sum_v(\Dop T)_v^n\sum_{v'\in V(v)}K_{vv'}^{1/2}(T^*_v-T^*_{v'})\\
    &=2\sum_v (\Dop T)_v^n (\Dop T)_v^*
    \end{split}
\end{equation}
The last expression is sign-definite in the limit of small time steps. The factor 2 appears because each $\Dop$ in (\ref{eq:BH equation}) leads to the same contribution. Thus, the sign-definite $\tilde{\chi}_{v}$ can be directly computed as $\tilde{\chi}_{v}=2(\Dop T)_v^n (\Dop T)_v^*$. The area-averaged results should be the same as if we were using the sign-indefinite form $\tilde{\chi}_v=\sum_{v'}(1/2)\tilde{\chi}_{vv'}$, even though they correspond to different spatial distributions (see further). Our manipulations are elementary by virtue of the selected edge-based operators and the fact that operators $\Dop$ are strictly symmetric. In the continuous case, the sign-definite form corresponds to the following chain (we suppress the 2 and let $K=1$ for brevity),
\begin{equation}
    \begin{split}
        &T\nabla\cdot\nabla (\nabla\cdot\nabla)T\\
        &=\nabla\cdot[T\nabla(\nabla\cdot\nabla T)]-\nabla T \cdot \nabla(\nabla\cdot\nabla T)\\
        &=\nabla\cdot[T\nabla(\nabla\cdot\nabla T)-(\nabla\cdot\nabla T)\nabla T]+(\nabla\cdot\nabla T)^2
    \end{split}
\end{equation}
The sign-indefinite form corresponds to the first equality. The flux related to it will be obtained similarly to (\ref{eq:vdf}). The second equality requires one more round of manipulations. A local discrete flux expression in this case does not necessarily exist. Once again, the flux of second moment  and the local DVD rate are ambiguous. However, one expects that flux divergence contribution will be efficiently damped on averaging, and both forms will lead to similar results.
\subsection{Advection}
Advection of tracer is commonly written in a flux  form as $\nabla\cdot(\mathbf{v}T)$. The advective flux of second moment can then be expressed in the following way,
\begin{equation}
\label{eq:AT}
\begin{split}
2T\nabla\cdot(\mathbf{v}T)&=\nabla\cdot(2T\mathbf{v}T)-2T\mathbf{v}\cdot\nabla T\\
&=\nabla\cdot(2T\mathbf{v}T)-\mathbf{v}\cdot\nabla T^2\\
&=\nabla\cdot(2T\mathbf{v}T)-\nabla\cdot(\mathbf{v}T^2)=\nabla\cdot(\mathbf{v}T^2)
\end{split}
\end{equation}
There are intermediate steps relying on the continuity $\nabla\cdot\mathbf{v}=0$, and it will be involved in derivation. To simplify the presentation we consider a regular 2D geometry with horizontal direction $x$ and vertical direction $z$. Advection of tracer $T$ in a layer with thickness $h$ is described by,
\begin{equation}
\begin{split}
&A_{k,i}[h^{n+1}_{k,i}T^{n+1}_{k,i}-h^{n}_{k,i}T^{n}_{k,i}]/\Delta t=\\
&-(U_{k,i+1/2}\tilde{T}_{k,i+1/2}-U_{k,i-1/2}\tilde{T}_{k,i-1/2}+[(W\tilde{T})|^t_b]_{k,i})
\end{split}
\label{eq:Tlayer}
\end{equation}
Here $t$ and $b$ denote the top and bottom faces of scalar control volume. They correspond to $k-1/2$ and $k+1/2$, with $k$ the vertical index increasing downward. $U=uh$ is the horizontal transport, $u$ the horizontal velocity, $W=wA$ the transport through layer interfaces, and $\tilde{T}$ the face value of tracer used to compute flux. One or both indices will be omitted if unambiguous. The quantity $A$ is the cell length in $x$ direction for 2D (it will become the horizontal area in 3D).  Simultaneously we have the thickness equation,
\begin{equation}
\begin{split}
 &A_{k,i}[h^{n+1}_{k,i}-h^{n}_{k,i}]/\Delta t=\\
&-(U_{k,i+1/2}-U_{k,i-1/2}+(W|^t_b)_{k,i})
\end{split}
\label{eq:thickness layer}
\end{equation}
We multiply the tracer equation (\ref{eq:Tlayer}) with $2T^*=T^{n+1}+T^n$, and group the terms on the left hand side,
\begin{equation}\label{eq:hT2}
\begin{split}
&A_{k,i}h^{n+1}_{k,i}(T^{n+1}_{k,i})^2 \\
&-A_{k,i}[h_{k,i}^n(T^n_{k,i})^2-T^n_{k,i}T^{n+1}_{k,i}(h^{n+1}_{k,i}-h^n_{k,i})]=\\
&-2T^*_{k,i}\Delta t(\mathbf{U}_{k,i+1/2}\tilde{T}_{k,i+1/2}-\mathbf{U}_{k,i-1/2}\tilde{T}_{k,i-1/2})\\
&-2T^*_{k,i}\Delta t[(W\tilde{T})|^t_b]_{k,i}
\end{split}
\end{equation}
The contribution due to layer motion is removed by using the thickness equation (\ref{eq:thickness layer}),
which leads for cell $k,i$ to,
\begin{equation}\label{eq:hT3}
\begin{split}
 &A\left[h^{n+1}(T^{n+1})^2 - h^n(T^n)^2\right]/\Delta t=\\
 &-U_{i+1/2}(2T^*\tilde{T}_{i+1/2}-T^nT^{n+1})\\
 &+U_{i-1/2}(2T^*\tilde{T}_{i-1/2}-T^nT^{n+1})\\
 &-[W(2T^*\tilde{T}-T^nT^{n+1})]|^t_b
\end{split}
\end{equation}
The advective terms here resemble the form of the third equality in (\ref{eq:AT}). As in the case of diffusion, we consider the contributions related to a face. We limit ourselves to the transport $U_{k,i+1/2}$. This transport is changing the second moment by,
$$
-U_{k,i+1/2}(2T^*_{k,i}\tilde{T}_{k,i+1/2}-T^n_{k,i}T^{n+1}_{k,i})
$$
in cell $k,i$ and by,
$$
U_{k,i+1/2}(2T^*_{k,i+1}\tilde{T}_{k,i+1/2}-T^n_{k,i+1}T^{n+1}_{k,i+1})
$$
in cell $k,i+1$. The total change in the second moment due to $U_{k,i+1/2}$ is,
\begin{equation}\begin{split}
\label{eq:chiA}
&\chi_{k,i+1/2}=
2U_{k,i+1/2}\tilde{T}_{k,i+1/2}(T^*_{k,i}-T^*_{k,i+1})\\
&-U_{k,i+1/2}(T^n_{k,i}T^{n+1}_{k,i}-T^n_{k,i+1}T^{n+1}_{k,i+1})
\end{split}\end{equation}
It is associated with the face $k, i+1/2$. Similar to the case of diffusion, we partition it equally between the two adjacent cells. This is equivalent to taking the flux of second moment,
\begin{equation}
\label{eq:hT4}
\begin{split}
&F_{k,i+1/2}=U_{k,i+1/2}\tilde{T}_{k,i+1/2}(T^*_{k,i}+T^*_{k,i+1})\\
&-\frac{1}{2}U_{k,i+1/2}\left(T^n_{k,i}T^{n+1}_{k,i}+T^n_{k,i+1}T^{n+1}_{k,i+1}\right)
\end{split}
\end{equation}
This flux differs from the advective flux in \citet{klingbeil2014quantification}, $U_{k,i+1/2}\tilde{T}_{k,i+1/2}^2$ (although both presents an approximation of the continuous flux of second moment). We delay their discussion to the following section.
We stress that the flux in (\ref{eq:hT4}) only determines how $\chi_{k,i+1/2}$ is partitioned between cells $k,i$ and $k,i+1$. The expression for $\chi_{k,i+1/2}$ can be modified to a slightly different form under assumption that $\Delta t$ is small. In this case $T^n_{k,i}T^{n+1}_{k,i}=(T^*_{k,i})^2+{\cal O}(\Delta t^2)$, and we get,
\begin{equation}
\label{eq:hT5}
\begin{split}
&\chi_{k,i+1/2}\approx\\
&2U_{k,i+1/2}(\tilde{T}_{k,i+1/2}-T^*_{k,i+1/2})(T^*_{k,i}-T^*_{k,i+1})
\end{split}
\end{equation}
where $T^*_{k,i+1/2}=\frac{1}{2}(T^*_{k,i}+T^*_{k,i+1})$. This approximate form illustrates that the DVD rate is related to deviation of $\tilde{T}$ from the central in time and space second-order estimate $T^*_{k,i+1/2}$. Indeed, if $\tilde{T}=T^*$, as in central in space Crank--Nicholson method, $\chi_{k,i+1/2}=0$. If $\tilde{T}$ is given by the first-order upwind method,
$\tilde{T}_{k,i+1/2}-T^*_{k,i+1/2}\approx \frac{1}{2}(T^*_{k,i} -T^*_{k,i+1})$ in the limit of small $\Delta t$, which corresponds to dissipation. However, there are issues for advection schemes of third or higher order, to be discussed further.
\subsection{Caveats}\label{caveats}
The concept of local DVD rate encounters certain difficulties. Take, for example, $T=\exp(-x^2)$ and $K=1$. For the harmonic diffusion we obtain $\chi=8x^2\exp(-2x^2)$. Dissipation occurs where the gradient of tracer is largest. For a biharmonic diffusive operator, the sign-indefinite form gives  $\chi=8(6x^2-4x^4)\exp(-2x^2)$, but the sign-definite one gives $\chi=8(2x^2-1)^2\exp(-2x^2)$. These forms predict different distributions in space. Since the biharmonic operator is commonly chosen from numerical considerations, there is limited physical motivation and the choice for the form of $\chi$ is not necessarily obvious. If we are willing to maintain similarity with harmonic diffusion, we may prefer the sign-indefinite form, but then the interpretation of the result as diffusion and anti-diffusion becomes just the result of this convention, i.e. it is ambiguous. If one will smooth $\chi$ with a broader kernel than the Gaussian in $T$, both forms will approximately agree, but the result will be just $((T^{n+1})^2-(T^n)^2)/\Delta t$ if diffusion is the only process. Presence of advection is the reason why this latter estimate is insufficient.\\

Advection introduces one more level of complications. The main difficulty lies in the elimination of advective flux of second moment from $\chi$. However, the numerical phase and group velocities deviate from the flow velocity depending on both temporal and spatial discretizations, and the elimination of advective transport cannot be complete. The form proposed here, if written for contributions from one direction, would imply for small $\Delta t$,
$$
\tilde{\chi}_i\approx U_{i+1/2}(\tilde{T}_{i+1/2}-T^*_{i+1/2})(T^*_i-T^*_{i+1})+
$$
$$
U_{i-1/2}(\tilde{T}_{i-1/2}-T^*_{i-1/2})(T^*_{i-1}-T^*_{i})
$$
The flux proposed in \citet{klingbeil2014quantification} will lead in the same approximation to a closed form,
$$\tilde{\chi}_i^{*}\approx -U_{i+1/2}(\tilde{T}_{i+1/2}-T^*_{i})^2+
$$
$$
U_{i-1/2}(\tilde{T}_{i-1/2}-T^*_{i})^2
$$
Note that here $\chi^*$ simply indicates it being obtained from \citet{klingbeil2014quantification} and does not imply the same meaning as $T^*$. In a 1D case with $U_{i+1/2}=U_{i-1/2}=u$ both expressions can be further simplified. We consider the case of the third-order upwind scheme, assuming that $u>0$, and that we are second-order accurate in time. In this case, $\tilde{T}_{i+1/2}=(1/2)(T^*_i+T^*_{i+1})-(1/6)(T^*_{i-1}-2T^*_i+T^*_{i+1})=T^*_{i+1/2}-D_i/6$. It is well known that this scheme has a truncation error in the form of biharmonic dissipation, but this is not explicitly seen in the expression for $\chi_{k,i+1/2}$.
Performing simple manipulations we find that,
\begin{equation}
    \begin{split}
    \tilde{\chi}_i\approx& \frac{u}{6}(2\partial_xT^*\partial_{xx}T^*-\Delta x\partial_xT^*\partial_{xxx}T^*)(\Delta x)^3\\
    \tilde{\chi}_i^*\approx& \frac{u}{6}(-\partial_xT^*\partial_{xx}T^*-\Delta x\partial_xT^*\partial_{xxx}T^*)(\Delta x)^3
    \end{split}
    \label{eq:small_dt}
\end{equation}
The second term in both expressions comes from the biharmonic dissipation $(|u|/12)(\Delta x)^3\partial_{xxxx}T$ of the third-order upwind advection scheme, and we see that it appears in both cases in a sign-indefinite form. However, in both cases there is an order larger term that corresponds to dispersive errors, which is not eliminated by the procedure proposed here nor by the procedure of \citet{klingbeil2014quantification}. This is not due to the dispersive truncation error of the third-order scheme, which has a higher order than the dissipation, it is the error entirely related to incomplete elimination of the flux divergence.  Note that the signs of dispersive terms in (\ref{eq:small_dt}) are different, and that the amplitude is twice larger for the method proposed here. One may try to use this fact (see below) but in real applications there are some contributions from temporal approximation and varying $u$, and the cancellation will not work.\\

One may readily find that in the case of fourth-order scheme, when $\tilde{T}_{i+1/2}=T^*_{i+1/2}-D_i/12-D_{i+1}/12$, the dissipative term disappears, but the dispersive term remains unchanged. In the case of second-order scheme, the method proposed here returns zero, but $\tilde{\chi}_i^*\approx -(u/2)\partial_xT\partial_{xx}T(\Delta_x)^3$, i.e. the dispersive term increases three times. Thus the DVD rate computed for advection contains a spurious term which is dispersive, i.e. $\chi$ contains some flux divergence. Simple definitions of $\chi$ or equivalently, fluxes here and in \citet{klingbeil2014quantification} are insufficient to fully eliminate dispersive contributions. The computed DVD rate cannot be interpreted in the sense of dissipation and anti-dissipation, i.e. local DVD rate. They need to be coarse-grained in order to have such sense. It is expected that coarse-graining will eliminate the dispersive error because $\partial_xT\partial_{xx}T=(1/2)\partial_x (\partial_{x}T)^2$.

\section{Experiments}
\label{sec:exp}
The method of this paper as well as that of \citet{klingbeil2014quantification} were implemented in the Finite-volumE Sea ice–Ocean Model (FESOM2) \citet{danilov2017finite}. The simulations reported here use a channel test case by \citet{soufflet2016effective}. Its basin is 2000 km long (North-South), 500 km wide (East-West) and 4 km deep, with periodic boundary conditions in zonal direction. Two triangular meshes with resolutions 10 km (40 vertical layers) and 5 km (60 vertical layers) have been used. Restart files were prepared, and simulations to diagnose the DVD were run from restarts with no external temperature forcing for simplicity. Physical mixing is due to background vertical diffusion of $10^{-5}$ m$^2$/s. FESOM uses longitude and latitude as coordinates, so the basin is approximately $18^\circ$ by $4.5^\circ$ degrees. Cosine of latitude is set to one to ensure flat geometry. \\

Several advection schemes were tested. They include the MUSCL-type scheme referred to as GE (gradient estimate), quadratic reconstruction (QR) scheme (see \citet{danilov2017finite}), central second-order scheme (C2) and a compact (CO) scheme (\citet{smolentseva2020comparison}) for horizontal advection. The compact scheme is so called because of analogy with one dimensional compact scheme, but this analogy persists only for uniform horizontal triangular meshes. The GE, QR and CO schemes are implemented in such a way that they mix third-order upwind and fourth-order central fluxes, and a numerical order parameter varying between 0 and 1 describes the share of the fourth order. Purely fourth-order schemes as well as second-order central scheme are stabilized with biharmonic horizontal diffusion. The biharmonic diffusivity coefficient is flow dependent in a Smagorinsky-like form and the proportionality coefficient is experimentally selected such that there is no apparent noise in the temperature field.\\

The vertical advection scheme is fourth order and is unchanged in our simulations. Since linear equation of state is used with temperature being the only tracer, the diagnosed DVD characterizes diapycnal mixing. Its physical part is due to background vertical mixing, while its numerical part is due to either advection or horizontal diffusion used for stabilizing advection. Because of the absence of surface forcing the physical vertical mixing is on the weak side in this setup. Our main intention is to learn about the level of numerical mixing and its spatial distribution. All measured DVD rates are in units $\textrm{T}^2/\textrm{s}$ implying tracer squared per second, which in this case is $^{\circ}\textrm{C}^2/\textrm{s}$. Also, $\tau=12$ mins for all experiments unless mentioned otherwise.

\subsection{Nomenclature}
To represent results consistently we first define terminology. $\chi$ will represent the total DVD rate per finite volume cell as obtained using the method discussed in this paper and $\chi^*$ will represent the total DVD rate obtained using the approach of \citet{klingbeil2014quantification}. Since we aim to decompose the total DVD rate into its constituents, we define some additional subscripts. `a' will denote advection (i.e., $\chi_a$ denotes DVD rate due to just advection as obtained using the method described in this paper). Similarly, subscript `d' will denote diffusion. We also introduce a second group of subscripts, `h' and `v' denoting horizontal and vertical directions respectively. We will use a bar to denote time averaging, $\overline{A}=(n\tau)^{-1}\sum_n A$, where $\tau$ is a single time step, and $n$ the number of time steps used for averaging. Finally, it will be reminded that in the context of this paper, numerical mixing is any mixing that cannot be attributed to a physical parameterization, i.e., $\chi_{\textrm{num}}=\chi_{\textrm{ah}}+\chi_{\textrm{av}}+\chi_{\textrm{dh}}$.

\subsection{Flux divergence uncertainty}\label{sec:fluxes}
We first illustrate the result (\ref{eq:small_dt}) comparing $\chi^*_{\textrm{a}}$ computed according to \citet{klingbeil2014quantification} and $\chi_{\textrm{a}}$ of this paper. They differ by the flux divergence.
\begin{figure}[h]
\noindent
\includegraphics[width=\columnwidth]{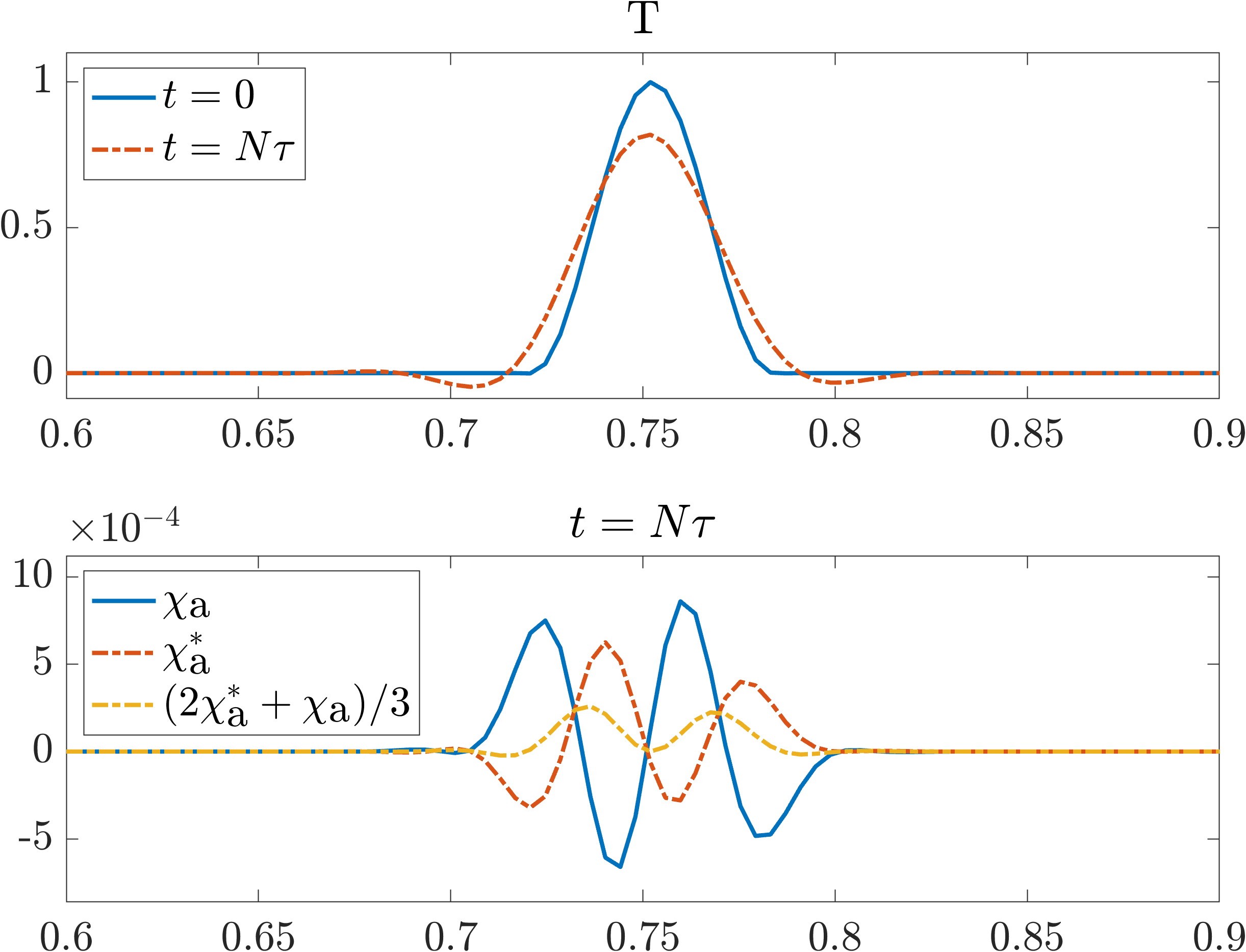}
\caption{Distribution of DVD rates due to 1D uniform advection for the third-order GE scheme. Top panel: Initial and final pulse shapes. Bottom panel: Final $\chi_{\textrm{a}}$ and $\chi^*_{\textrm{a}}$. }
\label{fig:1D_case}
\end{figure}
An initial pulse $T=\cos(0.5\pi(x-x_0)/\sigma)^2$, $|x-x_0|<\sigma$ with $x_0=0.75$ and $\sigma=1/32$ is advected by the third-order GE scheme on a mesh with $\Delta x=1/256$, $x \in \{0,1]$ by velocity $u=1$ using periodic boundary condition. The top panel of Fig. \ref{fig:1D_case} shows the initial distribution and the distribution after one complete period, while the bottom panel shows ${\chi}_{\textrm{a}}$ and ${\chi}^*_{\textrm{a}}$ diagnosed at the final step. Both behave in agreement with equation (\ref{eq:small_dt}). For this third-order case, the diagnosed DVD rates are dominated by the contributions from incompletely eliminated flux divergence (first terms in (\ref{eq:small_dt})). They are of opposite signs with the peaks of ${\chi}_{\textrm{a}}$ being larger. The dissipative part is revealed by  taking $(2\chi^*_{\textrm{a}}+\chi_{\textrm{a}})/3$, which is mostly positive, yet this can only be done for a uniform flow. Spatial averaging will damp the contribution from the flux divergence (not shown), but it will also smooth the dissipative part of the diagnosed DVD. We further investigate the effects of spatial and temporal averaging on this ambiguity of DVD estimate.\\

\begin{figure}[h]
\noindent
\includegraphics[width=\columnwidth]{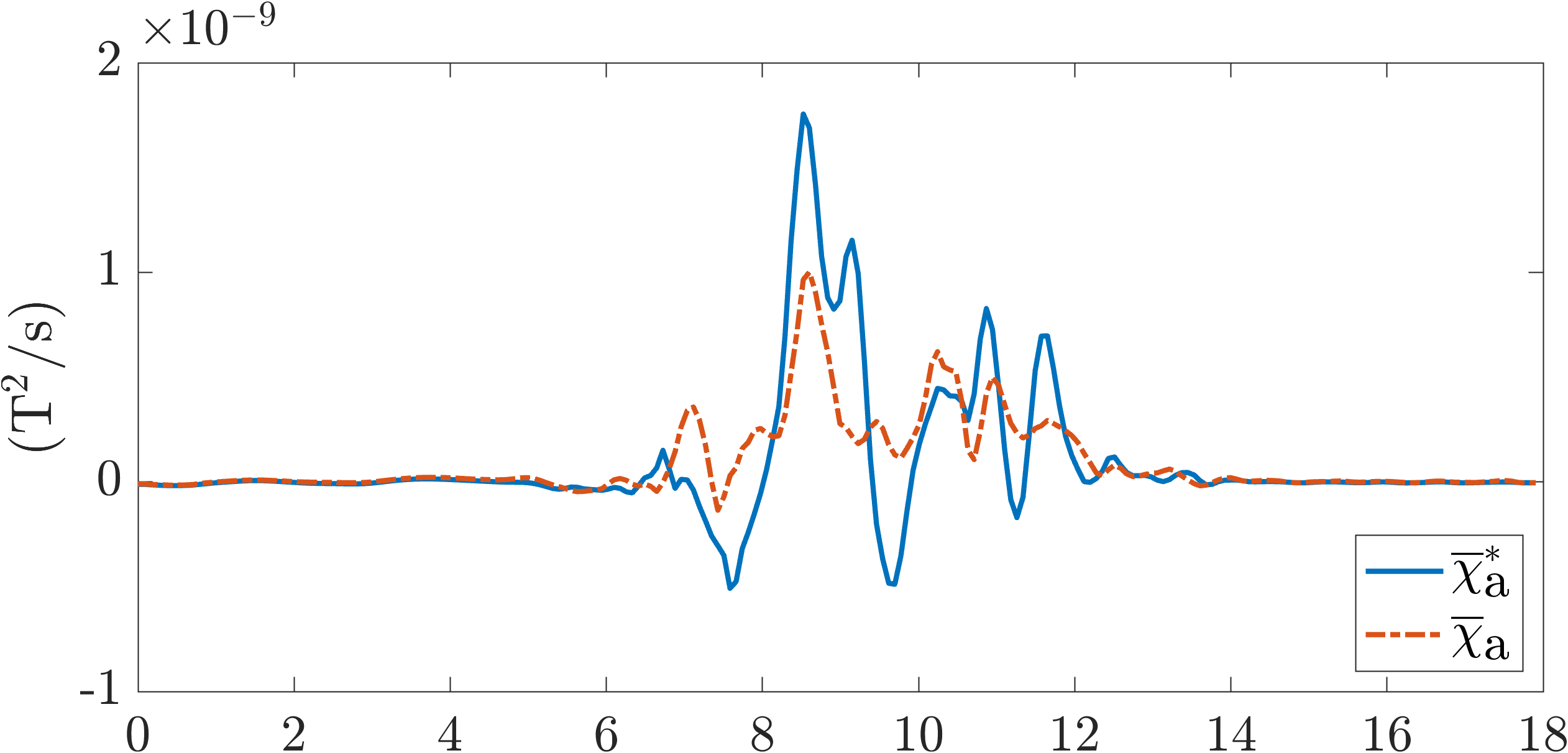}
\caption{Realizations of daily mean meridional profiles of DVD rates due to advection $\overline{\chi}_{\textrm{a}}$ and $\overline{\chi}^*_{\textrm{a}}$ on the 10 km mesh. The profiles are zonally and depth averaged. Fourth-order CO scheme is used (order parameter n = 1.0). The scheme still has residual dissipation leading to spurious mixing.}
\label{fig:ref1}
\end{figure}

\begin{figure}[h]
\noindent
\includegraphics[width=\columnwidth]{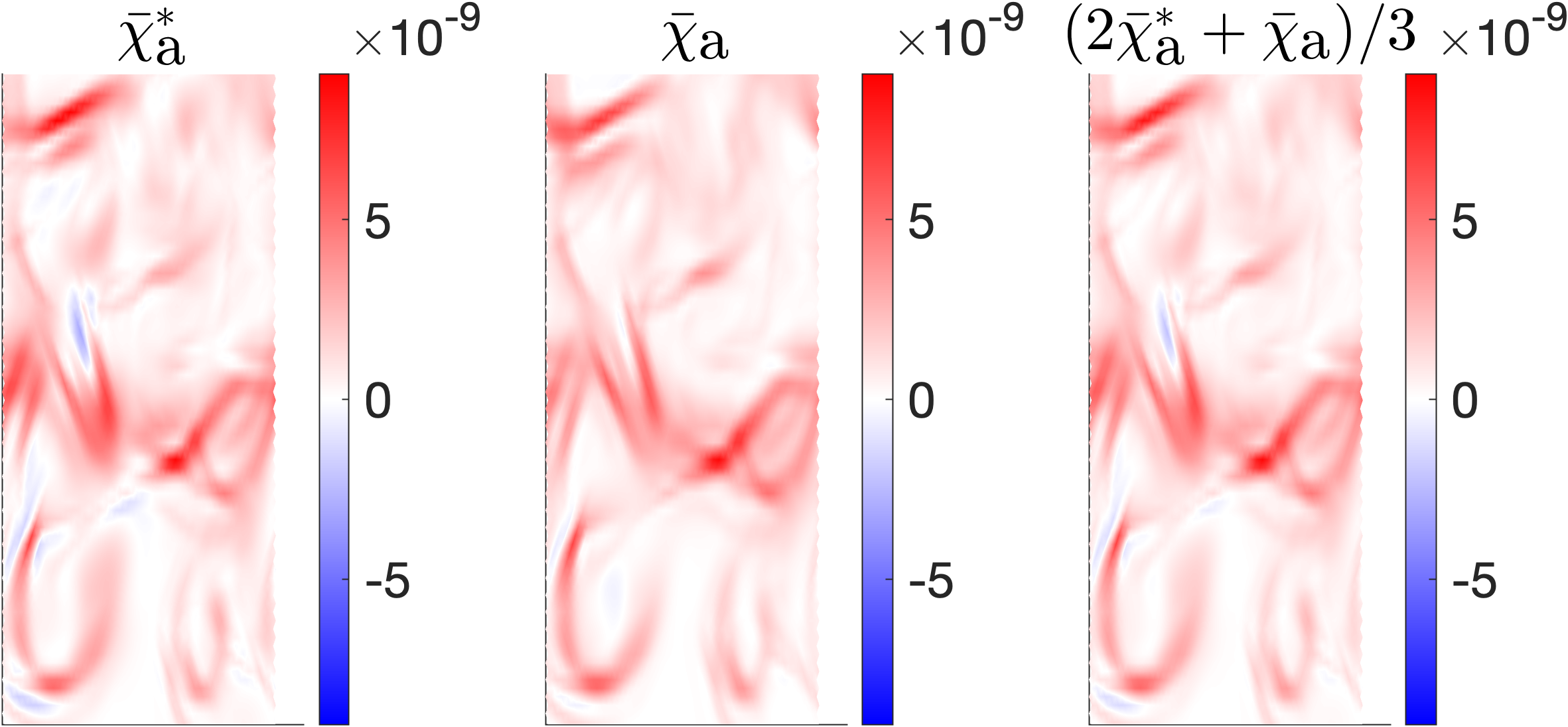}
\caption{Realizations of mean DVD rates due to advection $\overline{\chi}_{\textrm{a}}$ and $\overline{\chi}^*_{\textrm{a}}$ on the 10 km mesh between the meridional band 6 to 12 degrees. The distributions are depth averaged. The time averaging window is 2 months. Third-order CO scheme is used (order parameter n = 0.0).}
\label{fig:chi_3rd_comp}
\end{figure}

Figure \ref{fig:ref1} plots the DVD rates for a particular day in the channel setup. The plots here show meridional profiles, which are zonally and depth averaged. Although both  ${\chi}_{\textrm{a}}$ and ${\chi}^*_{\textrm{a}}$ agree in showing the strongest DVD rate at about 9 degrees and enhanced dissipation between 10 and 12 degrees, they still disagree in detail with ${\chi}^*_{\textrm{a}}$ characterized by stronger fluctuations. This shows the inability of such approaches to instantaneously estimate DVD rates reliably for high-order advection schemes with simple spatial averaging alone. In future, convolution with a kernel the size of a typical eddy may be more accurate than zonal averaging, ensuring better agreement between the two approaches. 
\\
 Time averaging alone also does not fully eliminate this discrepancy either as shown in figure \ref{fig:chi_3rd_comp}. Since these are realizations of a third-order scheme, the plots should be strictly positive (because of the built-in biharmonic dissipation term). We find this to be mostly true as they are predominantly positive, but all of them still contain some small negative patches. These patches do not imply anti-mixing but simply our inability to completely filter out the flux divergence from the DVD estimate. However, we do see that long temporal averaging can reduce this error to a satisfactory level, especially for $\overline{\chi}_{\textrm{a}}$ thereby making the spatial distribution close to expectation. We also see how the results from our 1D case with uniform velocity (figure \ref{fig:ref1}) do not translate to the general 3D case with non-uniform velocity (figure \ref{fig:chi_3rd_comp}). Here the best option appears to be $\overline{\chi}_{\textrm{a}}$ instead of $(2\overline{\chi}_{\textrm{a}}^*+\overline{\chi}_{\textrm{a}})/3$. However, it is not possible to claim this to be generally true.\\
 
 We take away from this section that the locality of such DVD-based approaches are likely not reliable without sufficient temporal or spatial averaging (ideally both).

\subsection{Decomposition}\label{sec:decomposition}
\begin{figure}[h]
\noindent
\includegraphics[width=\columnwidth]{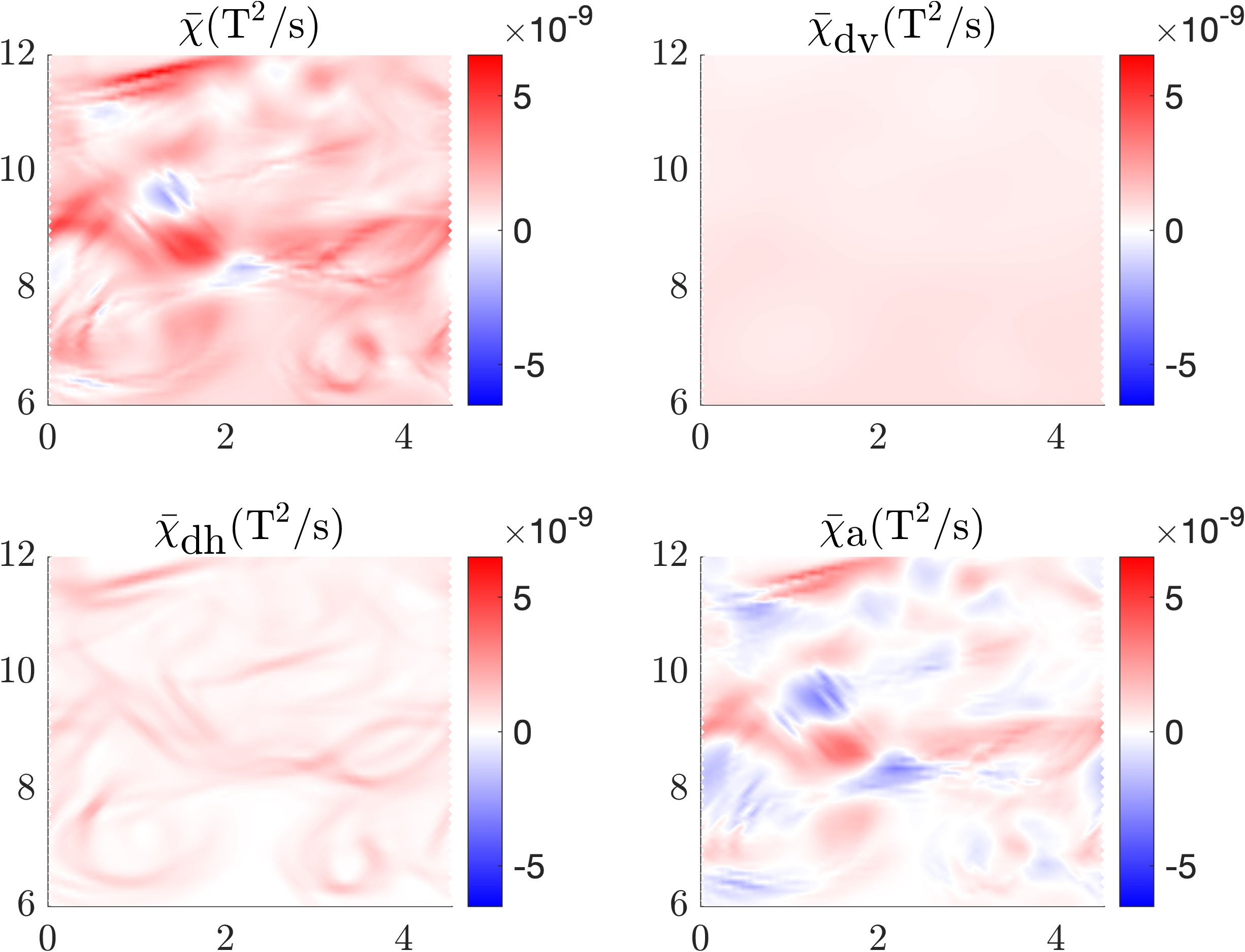}
\caption{Constituents of depth-averaged mean DVD rate $\overline{\chi}$  due to advection $\overline{\chi}_{\textrm{a}}$, horizontal diffusion $\overline{\chi}_{\textrm{dh}}$, and vertical diffusion $\overline{\chi}_{\textrm{dv}}$ on the 10 km mesh. The time averaging window is 2 months. Fourth-order CO scheme is used (order parameter $n=1.0$).}
\label{fig:decomposition_3months_CE}
\end{figure}

Our method allows one to split $\chi$ into its constituents as shown in figure \ref{fig:decomposition_3months_CE}. Simulations were carried out with the fourth-order CO scheme stabilized by explicit biharmonic horizontal diffusion. While conclusions based on strict local comparisons should not be drawn, broader inferences can still be made after sufficient time averaging (as explored in section \ref{sec:fluxes} previously and further verified below). The decomposition reveals how SDM ($\overline{\chi}_{\textrm{a}}+\overline{\chi}_{\textrm{dh}}$) can overpower physical mixing ($\overline{\chi}_{\textrm{dv}}$). From equation \ref{eq: Dver equation}, the physical diapycnal mixing (PDM) due to vertical diffusion will be sign-definite (for small time steps). Since vertical diffusivity is constant and small, PDM should be relatively unaffected by eddies in the channel setup depending only on the temperature profiles like the $\overline{\chi}_{\textrm{dv}}$ plot from figure \ref{fig:decomposition_3months_CE}.\\

For SDM due to biharmonic horizontal diffusion, it is not straightforward. Although net diffusion always results in variance decay, we use a sign-indefinite form (equation \ref{eq:BH equation}) which may locally include certain flux divergences. Nevertheless, it is expected to become sign-definite after sufficient time averaging like the $\overline{\chi}_{\textrm{dh}}$ plot of figure \ref{fig:decomposition_3months_CE}. $\overline{\chi}_{\textrm{dh}}$ also appears to be active only along the strong central eddy fronts. This is expected as biharmonic diffusivity in our implementation is defined to grow stronger with increasing velocity fluctuations. DVD due to advection meanwhile, has strong sign-indefinite tendencies (as discussed earlier). This should result in a patchy field like the $\overline{\chi}_{\textrm{a}}$ plot of figure \ref{fig:decomposition_3months_CE}. Even though we are dealing with the CO scheme of fourth order, it shows some residual dissipation, so the area-averaged $\overline{\chi}_{\textrm{a}}$ is positive.\\
\begin{figure}[h]
\noindent
\includegraphics[width=\columnwidth]{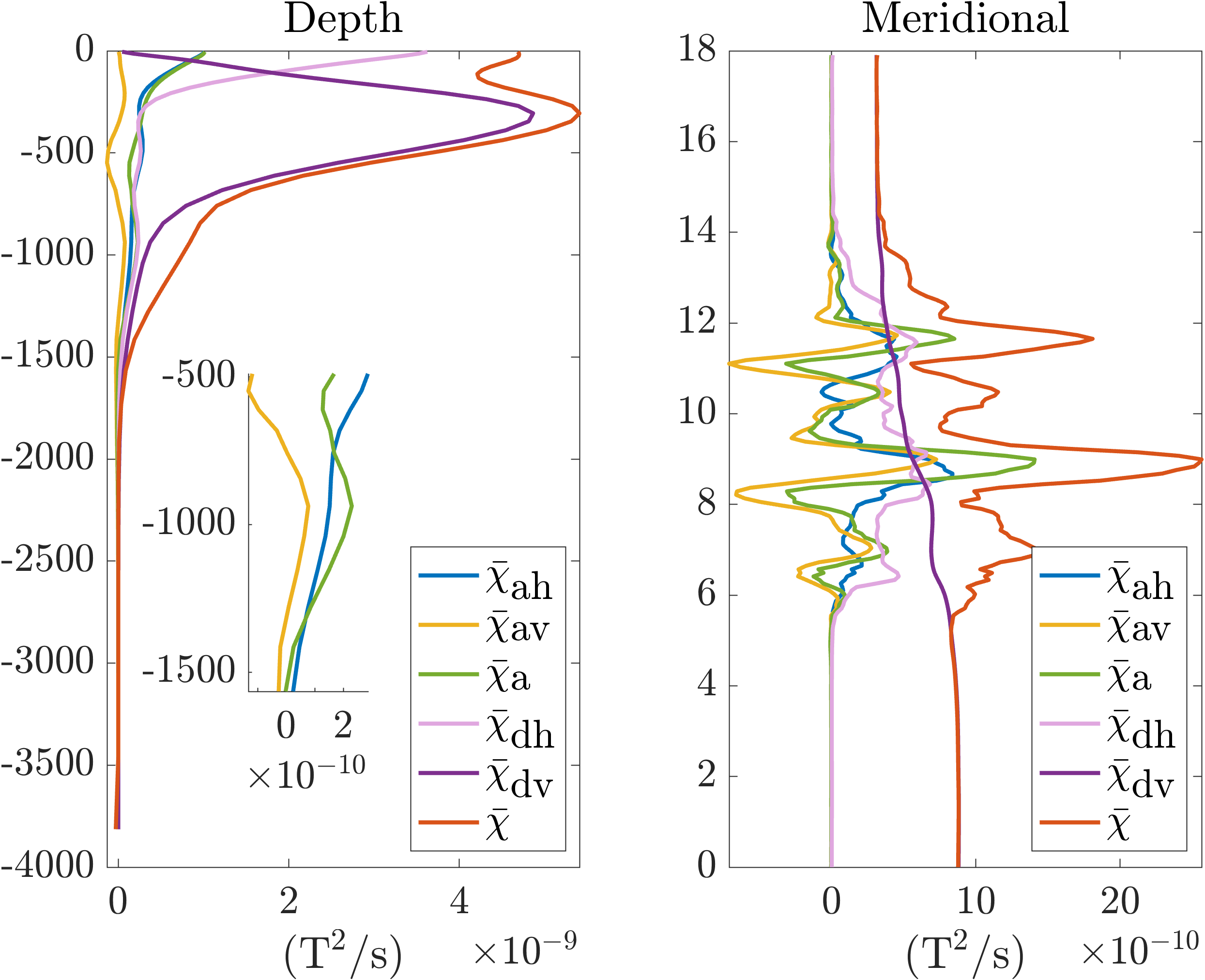}
\caption{Area-averaged depth profiles (left), zonally \& depth-averaged meridional profiles (right) of mean DVD rate $\overline{\chi}$ and its constituents - advection $\overline{\chi}_{\textrm{a}}$, horizontal diffusion $\overline{\chi}_{\textrm{dh}}$, and vertical diffusion $\overline{\chi}_{\textrm{dv}}$, on the 10 km mesh. $\overline{\chi}_{\textrm{a}}$ is decomposed into horizontal  $\overline{\chi}_{\textrm{ah}}$, and vertical  $\overline{\chi}_{\textrm{av}}$ components respectively. The time averaging window is 2 months. Fourth-order CO scheme is used (order parameter $n=1.0$).}
\label{fig:decomposition_mean_profiles_10km_CE}
\end{figure}\\
Figure \ref{fig:decomposition_mean_profiles_10km_CE} plots spatially averaged profiles which is expected to minimize most ambiguities caused by flux definitions in figure \ref{fig:decomposition_3months_CE}. It also demonstrates the ability to further decompose $\overline{\chi}_{\textrm{a}}$ into $\overline{\chi}_{\textrm{ah}}$ and $\overline{\chi}_{\textrm{av}}$ without operator splitting. \citet{soufflet2016effective} also provides reference vertical profiles of energy, buoyancy and temperature for this setup. Upon Comparing, $\overline{\chi}_{\textrm{dv}}$ is found to follow $\partial^2_{zz}T$ which is expected from equation \ref{eq: Dver equation}. For $\overline{\chi}_{\textrm{dh}}$, it is found to follow EKE. This is also expected because temperature gradients are large where eddies are present. Approximately, $\overline{\chi}_{\textrm{a}}$ too follows EKE. However, it has sub-surface (around $150$ m) and abyssal (around $1000$ m) peaks missing from the EKE profile. To understand these peaks, we inspect $\overline{\chi}_{\textrm{a}}$'s decomposition. The horizontal component $\overline{\chi}_{\textrm{ah}}$ is found mostly following EKE whereas the vertical component $\overline{\chi}_{\textrm{av}}$ is found following buoyancy flux instead. These peaks corresponded to depths associated with the largest conversions of APE to EKE. Indeed, one would then anticipate vertical advection to play a dominant role there. The behavior of the components $\overline{\chi}_{\textrm{a}},\overline{\chi}_{\textrm{dh}},\overline{\chi}_{\textrm{dv}}$ was thus found explainable and in agreement with expectations.

\subsection{Advection}
\begin{figure*}[h]
\noindent
\includegraphics[width=\textwidth]{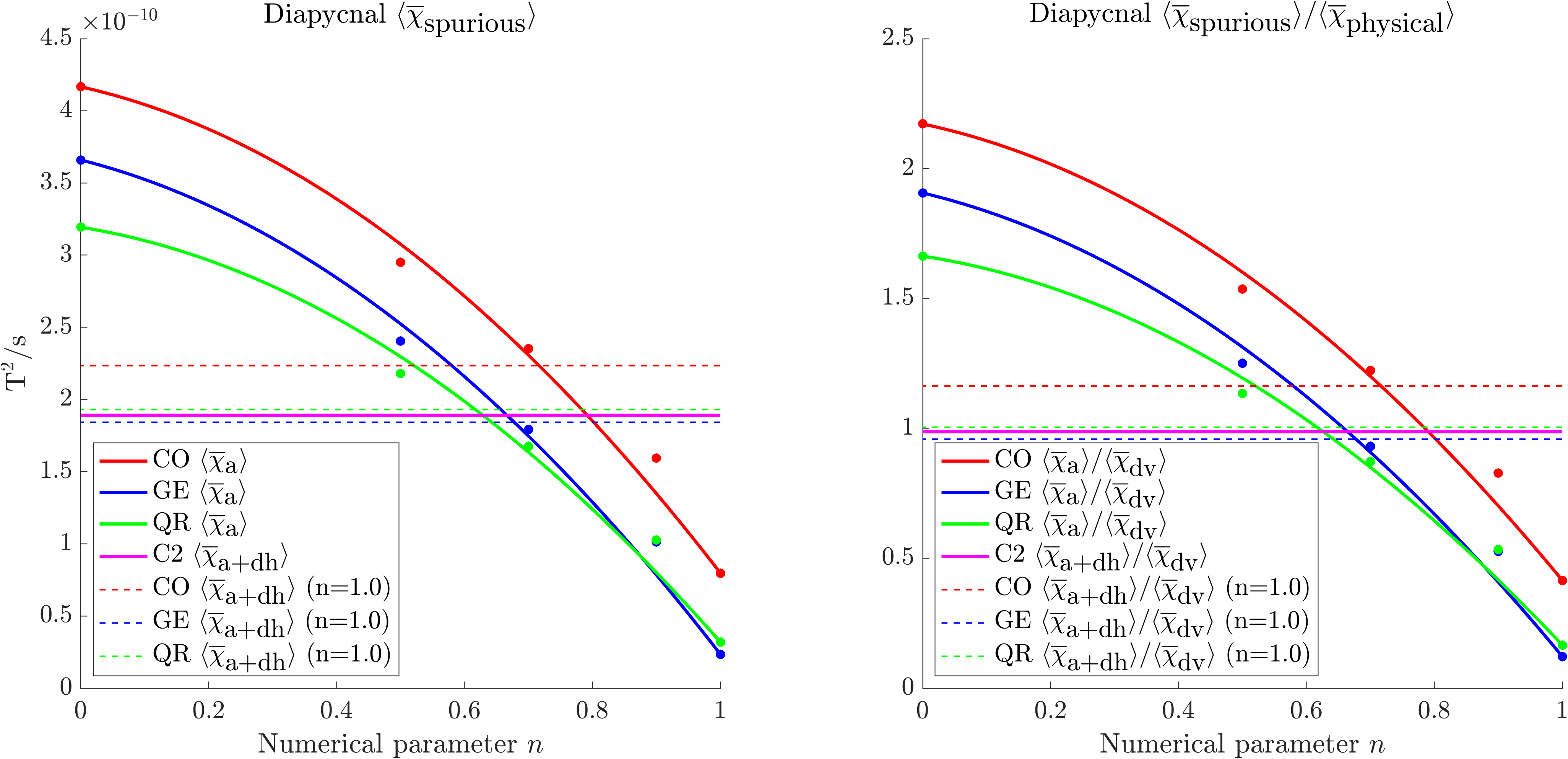}
\caption{Volume-averaged mean DVD rates leading to SDM (i.e., $\langle\overline{\chi}_{\textrm{a}}\rangle$, $\langle\overline{\chi}_{\textrm{dh}}\rangle$) for several advection schemes (defined earlier) as diagnosed on a 10 km triangular mesh. Here, the ``dots" represent measured $\langle\overline{\chi}_{\textrm{a}}\rangle$ data points, the solid lines are a second-order polynomial fit with 95\% confidence bound for the same, and ``dashed lines" are the measured values of $\langle\overline{\chi}_{\textrm{a+dh}}\rangle$ with numerical parameter $n=1.0$. The time averaging window is 2 months.}
\label{fig:advection_comp}
\end{figure*}

This section analyses SDM due to several advection schemes available in FESOM using the DVD technique described in this paper as shown in figure \ref{fig:advection_comp}. It plots second-order polynomial fits of experimentally obtained measurements for $n=1.0,0.9,0.7,0.5,0.0$\footnote{This was done to reduce the cost of calculating $\langle\overline{\chi}\rangle$ for every $n$.}. For the given resolution, every scheme appears to become more dissipative as their order is reduced from fourth to third. This is expected as a third-order upwind solution will have an additional biharmonic dissipation term. Within this idealized setup, the explicit biharmonic horizontal diffusion operator is also shown producing SDM similar to pure upwinding with $n\approx0.7$. Also observed is the closeness of SDM values as recorded between each scheme, i.e., one can be tweaked to achieve the SDM value of others. From the ratio of SDM to PDM, i.e., $\langle\overline{\chi}_{\textrm{spurious}}\rangle/\langle\overline{\chi}_{\textrm{physical}}\rangle$, the advection schemes are shown to produce less SDM (than PDM) only in numerically problematic regimes, i.e., with no separate biharmonic horizontal diffusion and $n\geq0.7$. While $n\geq0.7$ maintains stability for this idealized case, it can become unstable in more complex cases. FESOM is usually run with $n\leq0.5$ for global ocean simulations. Note that every scheme (for this idealized setup) produced more SDM than PDM i.e., $\langle\overline{\chi}_{\textrm{spurious}}\rangle\geq\langle\overline{\chi}_{\textrm{physical}}\rangle$ within their recommended operational limit $n\leq0.7$.\\

\subsection{Time stepping}
\begin{figure}[H]
\noindent
\includegraphics[width=\columnwidth]{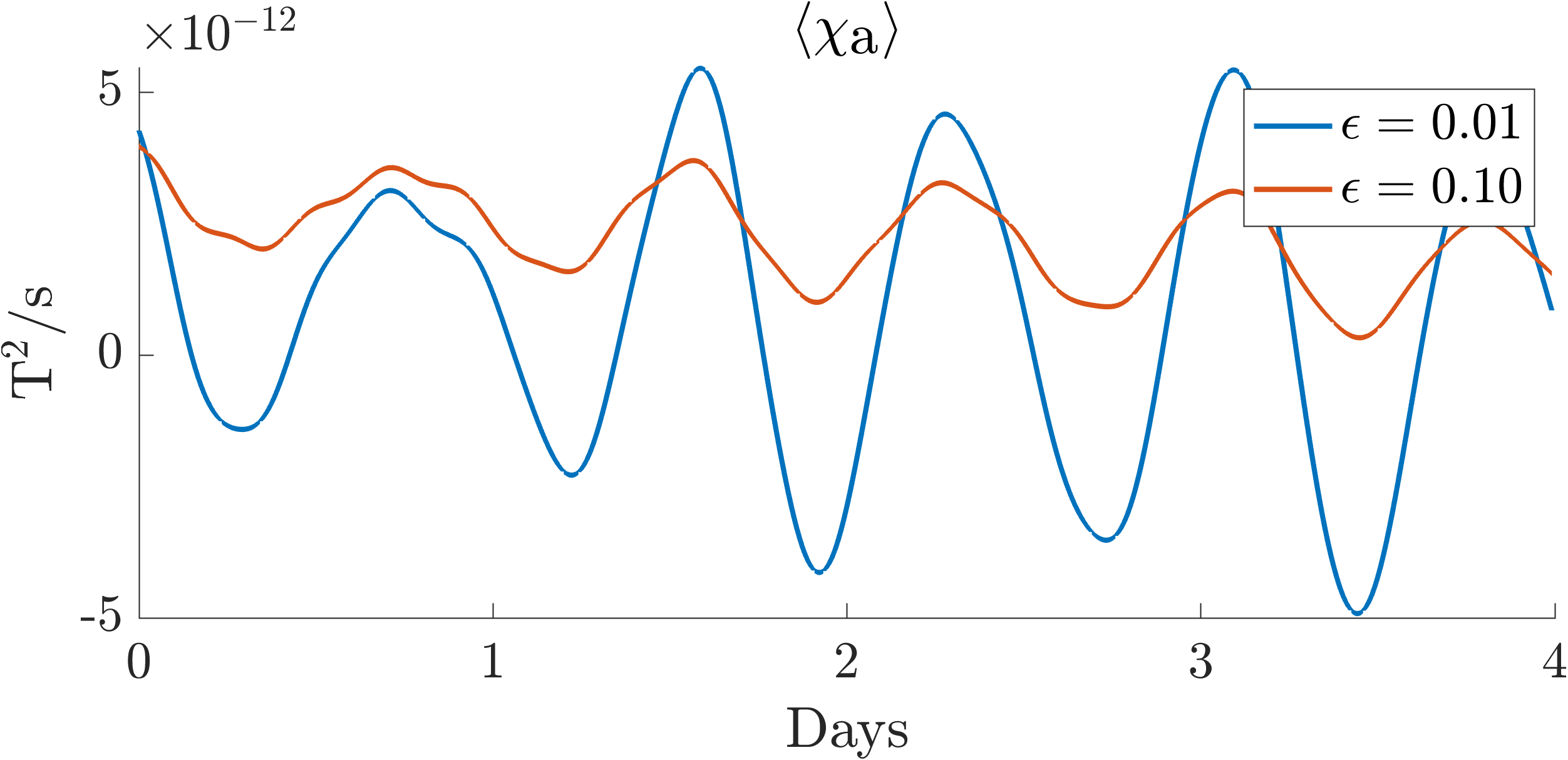}
\caption{Time evolution of volume-averaged DVD rate due to advection $\langle\chi_{\textrm{a}}\rangle$, on the 10 km mesh. $\epsilon$ is varied to introduce dissipation in the modified AB2 scheme for time interpolation while using Second-order C2 scheme for advection.}
\label{fig:epsilon}
\end{figure}

This section explores whether DVD rate $\chi_{\textrm{a}}$ is also impacted by dissipation from temporal interpolation. In FESOM the second-order Adams Bashforth (AB2) scheme of the form $q^{AB2}=(3/2+\epsilon)q^n-(1/2+\epsilon)q^{n-1}$ is used to interpolate tracer in time. Here $\epsilon$ is a small parameter $\leq0.1$ required for numerical stability. Figure \ref{fig:epsilon} plots the measured DVD rates $\chi_{\textrm{a}}$ for $\epsilon=0.01$ and $0.1$ using C2 scheme. Stability for the smallest $\epsilon$ is ensured by biharmonic diffusion. They show stronger damping of inertial oscillations for larger $\epsilon$ and a systematic increase in mean $\chi_{\textrm{a}}$ for larger $\epsilon$. This is expected because non-zero $\epsilon$ will introduce an additional dissipation term. Dissipation related to time stepping therefore worsens $\chi_{\textrm{a}}$ resulting in elevated values of SDM. However, in the test case considered, its contribution is rather weak compared to other contributions.\\

\subsection{Resolution}
\begin{figure}[H]
\noindent
\includegraphics[width=\columnwidth]{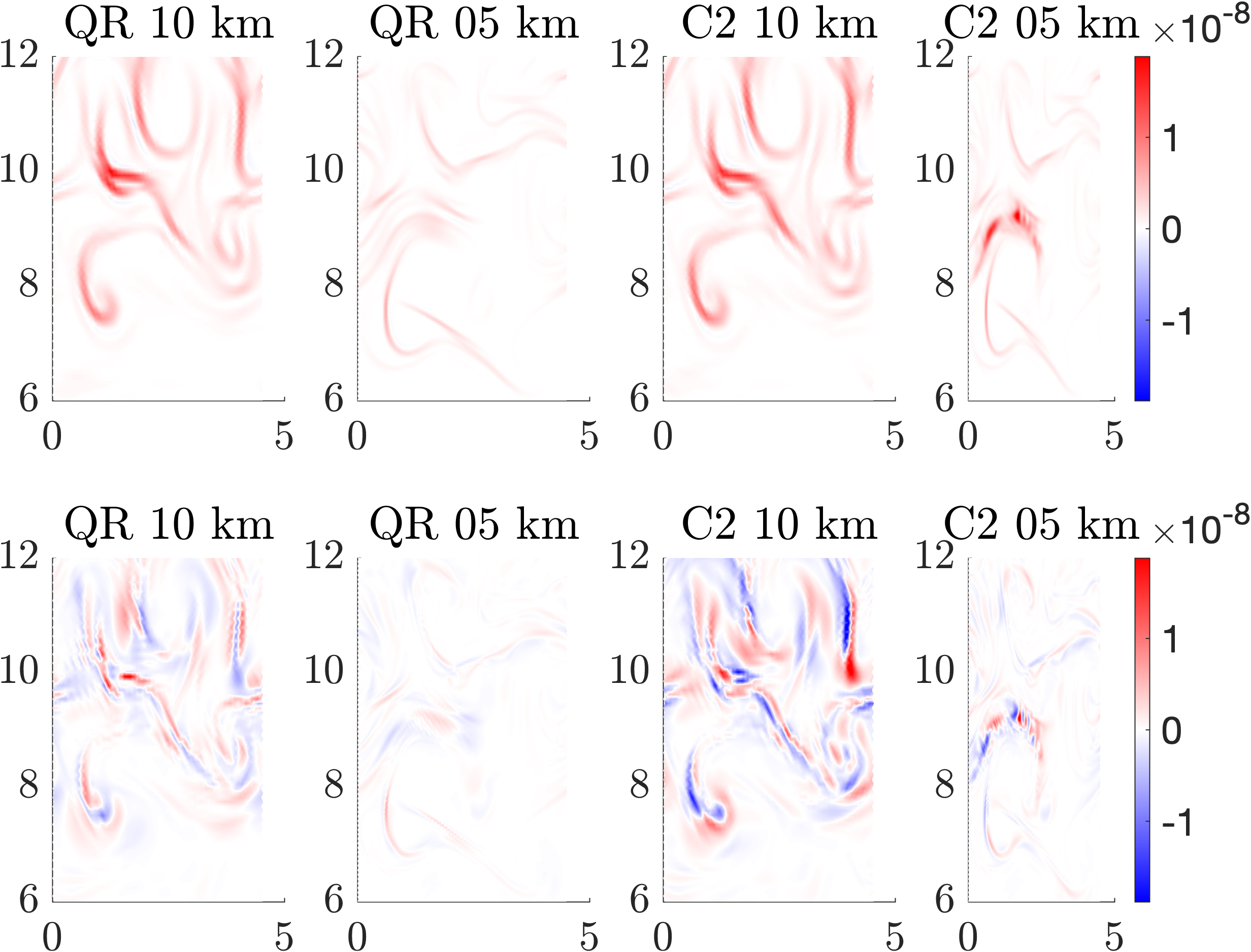}
\caption{DVD rates due to horizontal diffusion (top) and advection bottom obtained with C2 and QR schemes on meshes with resolution 10 and 5 km. The time averaging window is 1 month. The C2 scheme used is second-order and the QR scheme used is fourth-order (order parameter n = 1.0).}
\label{fig:ref3}
\end{figure}

\begin{figure}[H]
\noindent
\includegraphics[width=\columnwidth]{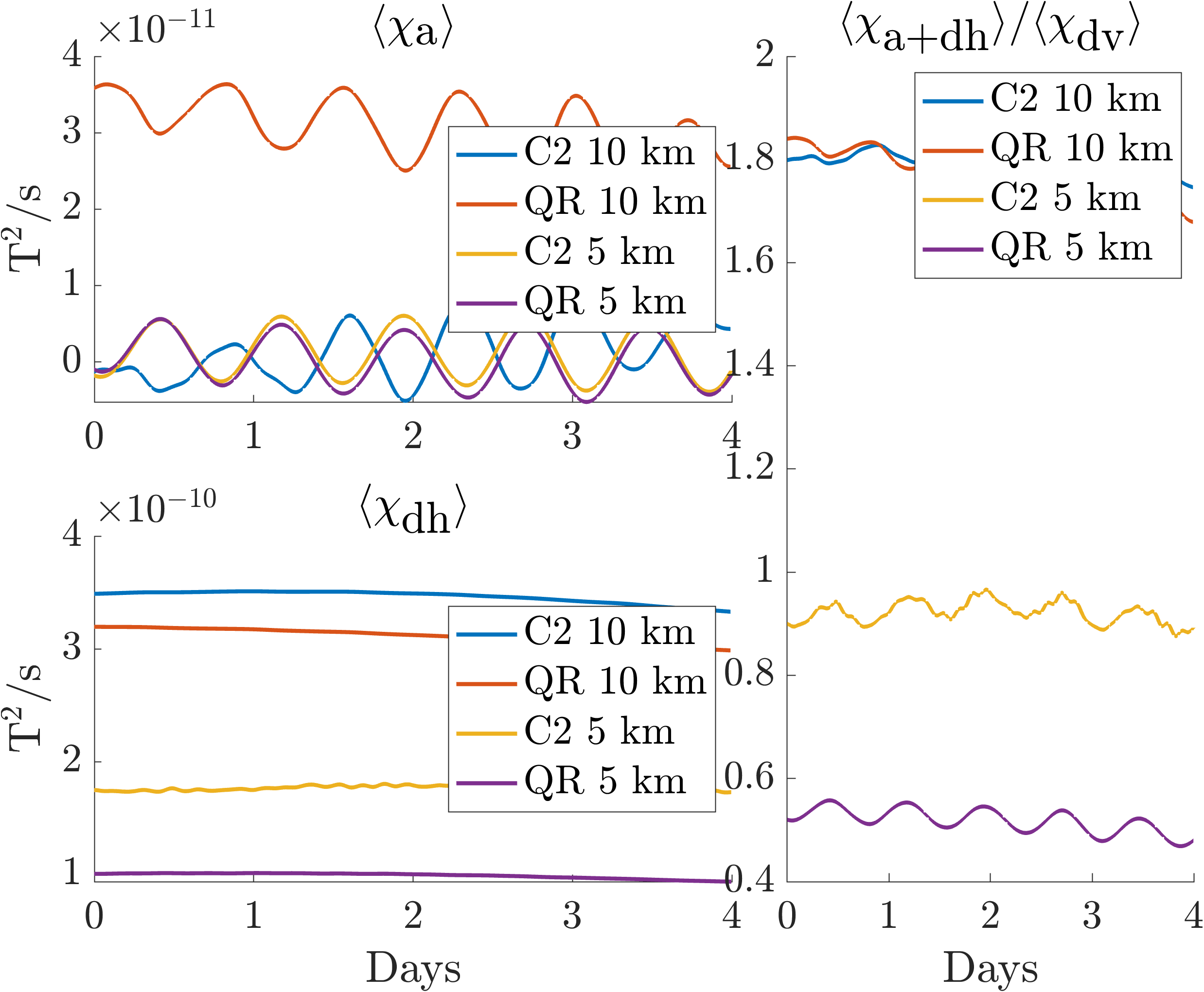}
\caption{Time evolution of volume-averaged DVD rates - spurious $\langle\chi_{\textrm{a+dh}}\rangle$, advection $\langle\chi_{\textrm{a}}\rangle$, and horizontal diffusion $\langle\chi_{\textrm{dh}}\rangle$ on the 10 km mesh. The (right) panel compares them to physical vertical diffusion $\langle\chi_{\textrm{dv}}\rangle$. These plots are for advection schemes - C2, and QR with order parameter $n=1.0$.}
\label{fig:resolution_abs}
\end{figure}

One may hope that mesh refinement will reduce SDM because both explicit horizontal diffusion and diffusion build in advection schemes scale with resolution as $l^3$ in our setup, while eddy kinetic energy increases very moderately for resolutions finer than 10 km. On the other hand, sharper tracer gradients might be simulated on a finer mesh, so that the answer is not obvious. We compare the DVD rates in the channel for two uniform triangular meshes with resolution 5 and 10 km (time steps 6 and 12 min respectively) using C2 and QR ($n=1$) schemes. These schemes have dispersive truncation errors, and spurious dissipation is mostly due to horizontal biharmonic diffusion. Figure \ref{fig:ref3} presents vertically integrated $\chi_\mathrm{dh}$ (top panels) and $\chi_\mathrm{a}$ (bottom panels) averaged over 1 month. It illustrates the reduction in both $\chi_\mathrm{dh}$ and $\chi_\mathrm{a}$ on mesh refinement. Also note that QR scheme leads to smaller values of $\chi_\mathrm{dh}$ and $\chi_\mathrm{a}$ than C2 scheme on the finer mesh, which agrees with this scheme being more accurate. Time evolution of basin-mean DVD rates due to advection and horizontal diffusion is plotted in Fig.  \ref{fig:resolution_abs}. The measured DVD rates for PDM, i.e., $\chi_{\textrm{dv}}$ (not shown) do not change with resolution, as they only depend on the vertical tracer gradient which is very similar on the two meshes. For the 10 km mesh, SDM  for both schemes is  higher than PDM (approximately 2$\times10^{-10}$ $^{\circ}\textrm{C}^2/\textrm{s}$). The situation reverses when resolution is increased to 5 km. As can be seen from Fig. \ref{fig:resolution_abs}, dissipation is mostly due to horizontal diffusion. Note that on the 10 km mesh, the QR scheme also contributes to dissipation (about 10\% of diffusive contribution) which is absent in the 5 km mesh. Thus, a more accurate scheme is definitively less dissipative only on a sufficiently fine mesh. On coarse meshes, their dissipation characteristics are no longer obvious. The same reasoning can also be applied to upwind schemes ($n<1$) as well. In that case, the advection schemes would report much higher SDM and will be its sole contributor.\\
\section{Discussions}\label{sec:disc}
Discrete tracer equation and its corresponding second moment form can be written symbolically at any location such that the LHS is the local variance decay, and the RHS are contributions of horizontal and vertical, advection and diffusion. Despite the seeming simplicity, DVD rate due to a particular process cannot be estimated by taking the respective term on RHS with a minus sign because such an estimate will include contributions of flux divergence. Contributions of diffusive fluxes are expected to be localized in some neighbourhoods where dissipation is occurring and might be eliminated by some area averaging. In contrast, the advective fluxes can be large-scale and need to be eliminated directly. The related discrete fluxes are unknown because the discrete equation for second moment estimated from the equation for first moment does not have a flux form even if a flux form is used for the first moment. One can either propose a physically sound definition of fluxes, as done in \citet{burchard2008comparative}, \citet{klingbeil2014quantification}, or re-arrange the RHS terms in a way resembling manipulations in the continuous case, as is done in this paper. This re-arrangement is free of additional assumptions and is only a framework change from cells (control volumes) to their faces. Indeed, the contribution from a flux into cells on both sides of the face cancel if taken together, and one is left with sinks or sources  associated with the face, which is the DVD rate. It can be partitioned equally between the cells separated by this face. The partitioning defines implied fluxes of second moment.\\

These fluxes generally differ from those proposed by \citet{klingbeil2014quantification}, but agree in the limit of small time steps for diffusion, and in the representation of dissipation for advection. In case of harmonic diffusion, the $\chi$ obtained approximates continuous variance decay. If applied to the biharmonic diffusion, however, the procedure returns a sign-indefinite form of $\chi$. Although in practice some temporal averaging might be sufficient to make it sign-definite, this implies that flux divergences are not entirely eliminated. Indeed, another round of identical transformations can be proposed if the biharmonic operator is a product of two identical symmetric harmonic operators, leading to a sign-definite form of $\chi$. This creates an ambiguity. Unfortunately, a precise form of build-in diffusion in high-order advection operators is not necessarily known to warrant additional manipulations. The implication is that the $\chi$ due to advection may still contain some divergence of non-compensated flux of second moment, as shown earlier. This is a drawback shared with \citet{klingbeil2014quantification}. The $\chi$ due to advection often contains negative patches. Interpreting them as anti-diffusion is not rigorous in a general case.\\

It must be acknowledged that the DVD rate is not a {\em local} quantity unless one is limited to harmonic diffusion and lower-order advection. It generally needs certain coarse-graining, i.e. averaging in space and time, although time averaging might be sufficient in some cases. The DVD framework proposed here implies slightly more work than needed to implement the analysis by \citet{klingbeil2014quantification} because it relies on $T^*$ which is available only when the time step $n+1$ is completed. One therefore needs to additionally store the fluxes of first moment and $T^n$. However, the computations are strictly consistent with the discrete tracer equations, and they more clearly reveal the source of ambiguity for the local interpretation of DVD rate. Both approaches can be used in practice subject to limitations mentioned above. Our framework also reveals how generalized directional decomposition of DVD rates may be achieved without operator splitting, which is not obvious from previous publications on the topic. In reality, we see that this approach can admit any arbitrary flux definition, including the ones from earlier publications.\\

Despite uncertainties caused by flux divergence, DVD can be used as a convenient tool to diagnose numerical mixing associated with advection or stabilizing factors (limiters or horizontal diffusion), which is the main `unknown'. This is provided sufficient averaging (space or time) is performed. In this paper, we do not explore such possibilities beyond simple zonal, depth, and mean profiles. It remains to be seen in the future if some more intricate averaging yields a better result. In a general case, in $z$- coordinate models the diagnosed DVD rate does not distinguish between isoneutral and dianeutral mixing, and a question arises whether the diagnostic can be extended to allow such a split. This is also a subject of future work.\\

A further central question is how to reduce the undesired effects of numerical mixing leading to SDM. According to \citet{Ilicak2012}, SDM is highly sensitive to the grid-scale Reynolds number, which has not been illustrated here. A particular question is on the combination of viscosity/upwinding and diffusion/upwinding which leads to smallest SDM and simultaneously does not suppress eddy motions. \citet{lemarie2012} proposed using biharmonic isoneutral diffusion instead of horizontal diffusion, in a combination with the fourth-order advection to reduce spurious effects of the third-order scheme. The biharmonic isoneutral operator needs stabilization in the form of vertical diffusion, which will be the only numerical dissipation if applied to the test case above. It remains to be seen to what an extent such measures or the introduction of $\tilde{z}$ vertical coordinate as done by \citet{Petersen2015} and \citet{megannztilde} will be helpful in the case of FESOM. The DVD rate diagnostics will be a helpful tool in all these cases.         

\section{Conclusion}\label{sec:conc}
This paper explored the use of Discrete Variance Decay (DVD) approach for diagnosing Spurious Mixing (SM) in ocean models. Its limitations were addressed and an alternative framework to \citet{klingbeil2014quantification} was proposed. Its performance was assessed and it was applied to several advection schemes in an idealized setup where SM was also the Spurious Diapycnal Mixing (SDM). To summarize,\\
\begin{itemize}
\item Closed-form expressions: Direct estimates for DVD rates due to diffusion and advection alongside their horizontal and vertical components were obtained. The source of ambiguity associated with second-moment fluxes and how it is unavoidable for simple flux definitions were also revealed. It was shown how this leads to uncertainty in localized predictions. The necessity for any DVD-based technique to be followed by some spatial or temporal averaging, ideally both, was further demonstrated.
\item Numerical mixing: SM was isolated and decomposed into respective constituents. For linear monovariate equation of state, SM became the SDM. Net SDM was found to be highly correlated with the distribution of eddy kinetic energy. The contribution from vertical advection was found to be relatively small and correlated with the distribution of buoyancy fluxes.
\item Advection schemes: Several advection schemes with varying orders of accuracy available in FESOM2 were tested. All schemes recorded similar levels of SDM which increased with the upwinding or explicit horizontal diffusion used. Throughout their recommended operational limit (i.e. for numerical parameter $n\leq0.7$), they were found to produce more SDM than Physical Diapycnal Mixing (PDM). This SDM could be as high as twice the background PDM for third-order schemes (i.e. $n=0$).
\end{itemize}
\section{Acknowledgements}
This paper is a contribution 
to  the Collaborative Research Centre TRR 181 ``Energy Transfers in Atmosphere and Ocean'' 
funded by the Deutsche Forschungsgemeinschaft (DFG, German Research Foundation) - Projektnummer 274762653. 
\bibliographystyle{elsarticle-num-names}
\bibliography{paper}







\end{document}